\documentclass[3p,times]{elsarticle}

\usepackage{ecrc}

\volume{00}

\firstpage{1}

\journalname{Nuclear Physics A}

\runauth{}

\jid{npa}

\jnltitlelogo{Nuclear Physics A}

\usepackage{graphicx}
\usepackage{graphics}
\usepackage{dcolumn}
\usepackage{mathrsfs}
\usepackage{amssymb}
\usepackage{amsmath}
\usepackage[pdftex]{hyperref}
\usepackage{natbib}
\usepackage{xspace}
\usepackage{slashed}
\usepackage{nicefrac}

\biboptions{square,comma,numbers,sort&compress}

\usepackage[figuresright]{rotating}

\def\del{\partial}

\long\def\comment#1{ }

\def\p{{\boldsymbol p}}

\def\0{{\boldsymbol 0}}
\def\q{{\boldsymbol  q}}

\def\k{{\boldsymbol k}}

\def\x{{\boldsymbol x}}
\def\y{{\boldsymbol y}}

\def\r{{\boldsymbol r}}

\newcommand{\beq}{\begin{eqnarray}}
\newcommand{\eeq}{\end{eqnarray}}
\newcommand{\be}{\begin{eqnarray*}}
\newcommand{\ee}{\end{eqnarray*}}
\newcommand{\bal}{\begin{align}}
\newcommand{\eal}{\end{align}}

\newcommand{\rmTr}{{\rm Tr}}

\newcommand{\nn}{\nonumber\\ }

\begin{document}

\begin{frontmatter}



\dochead{}

\title{Probabilistic picture of in-medium jet evolution}


\author[cea]{Yacine Mehtar-Tani}

\address[cea]{Institut de Physique Th\'eorique,
CEA Saclay, F-91191 Gif-sur-Yvette, France}

\begin{abstract}
We briefly discuss the recently developed probabilistic picture for in-medium jet evolution that is driven by independent multiple scatterings and branchings. These are controlled by the jet quenching parameter $\hat q$. In this framework, large radiative corrections to $p_\perp$-broadening of partons in the jet, enhanced by a double logarithm (DL) of the medium size $L$, are recovered. We argue that these non-local corrections are universal and can be reabsorbed in a renormalization of the jet quenching parameter without spoiling the probabilistic picture. As a consequence, we find that for large media, the mean radiative energy loss result scales as  $L^{2+\gamma}$, where the anomalous dimension $\gamma=2\sqrt{\alpha_sN_c/\pi}$. 
\end{abstract}


\end{frontmatter}


\section{The approximation of independent multiple scatterings and branchings}
\label{sec:Introduction}

In a large QCD medium, a high energy parton might undergo multiple interactions.  So long as their range is much smaller than the mean free path $\lambda$ of the parton in the medium they can be treated as independent from one another. The physics is then fully encoded in a single local interaction which at leading order in perturbative QCD  is given by the elastic cross-section $d\sigma_\text{el}/d^2\q\simeq g^4/\q^4 $, where $g\ll 1$ is the coupling constant and $\q\gg \Lambda_\text{QCD}$  is the transverse momentum exchanged in the scattering process. In this case, the so-called quenching parameter $\hat q$ corresponds to a (local) diffusion coefficient in transverse momentum space with respect to the direction of its large momentum component \cite{BDMPS}, that is, 

\beq\label{qhat}
\hat q\equiv \frac{d\langle k_\perp^2\rangle}{dt}=n\int\frac{d^2\q}{(2\pi)^2} \q^2 \frac{d^2\sigma_\text{el}}{d^2\q}\,,
\eeq
where $n$ is the density of medium color charges. This yields a typical transverse momentum broadening, for a high-energy particle, $\langle k_\perp^2\rangle_\text{typ}=\hat q L$, in a medium of size $L$. Note that, this approximation is justified in a weakly coupled quark-gluon plasma (QGP). Indeed, in that case the range of medium interactions, given by the inverse Debye mass $m_D\sim gT$, where $T$ is the temperature of the plasma, is parametrically much smaller than the mean-free-path $\lambda \sim (g^2T)^{-1}$. In terms of microscopic scales, $\hat q \simeq m_D^2/\lambda$.

In addition to $p_\perp$-broadening, multiple scatterings might induce fast radiations of soft gluons off the energetic parton with constant rate along the medium, hence suppressing its energy \cite{BDMPS,Zakharov}. It takes a coherence time $\tau=\omega/k_\perp^2$ for a gluon of energy $\omega$ to form. When it is formed, it has accumulated in the medium a typical transverse momentum $k^2_\perp \sim \hat q \tau$, via multiple scatterings. This yields the characteristic formation time $\tau_\text{form}(\omega)\equiv \sqrt{2\,\omega/\hat q}$. The leading-order (LO) gluon radiation probability, the so-called BDMPS spectrum, reads (for $\tau_\text{form}\ll L$)
\beq\label{BDMPS-1}
\omega\frac{ dN}{d\omega} \simeq\frac{\alpha_sC_R}{\pi} \frac{L}{\tau_\text{form}}\,,
\eeq
and yields a mean-energy loss $\langle\omega \rangle\sim \alpha_s C_R \,\hat q\, L^2$ where $C_R$ is the color charge of the emitter. This mechanism provides a natural explanation for the observed suppression of jets in heavy ion collisions compared to proton-proton collisions at RHIC and the LHC \cite{d'Enterria:2009am,Mehtar-Tani:2013pia}.  When $\tau_\text{form}\lesssim \alpha_s L$, the BDMPS spectrum can no longer be identified with a probability, and higher order corrections become non-negligible. In the limit of large media, these corrections can be resummed \cite{BDIM}. Multiple gluon emissions can then be treated as independent when the formation time is much smaller than the radiative mean free path, i.e., the typical time between two successive emissions, $\lambda_\text{rad}\equiv \tau_\text{form}(\omega_\text{BH})/\alpha_s\simeq \lambda/\alpha_s$ and $\omega_\text{BH}\sim m_D^2\lambda$ is the Bether-Heitler infrared cut-off (we have used to fact that $\hat q\sim m_D^2/\lambda$).  We observe that this is indeed realized in the small coupling limit when $\omega \ll  \omega_\text{BH}/\alpha_s^2$. Combining the two inequalities we get the condition for the validity of multiple independent interaction approximation: $  \tau_\text{form}\ll \lambda_\text{rad}\ll L\, \lesssim\, \lambda /\alpha_s^2$.
Not only gluon radiation can be treated as quasi-local compared to the size $L$ of the system but further evolutions of the off spring gluons are independent from on another \cite{BDIM}. This is the basis of in-medium jet evolution. In this framework, both the local scattering and branching rates depend on medium properties solely via the quenching parameter $\hat q$.  

Recently, radiative corrections to transverse momentum broadening have been considered \cite{Wu:2011kc,Liou:2013qya}. Potentially large corrections, enhanced by a double logarithm of the length of the medium $L$, were found. In light of this result, one can also expect large radiative corrections to radiative energy loss and other in-medium jet observables as well. Moreover, the large non-local radiative corrections can be expected to spoil the classical picture of independent multiple interactions and branchings.  We argue that the double logarithmic corrections are universal and can be absorbed in a renormalization of the jet quenching parameter. Hence, the probabilistic approximation remains valid \cite{Blaizot:2014bha}. 

\section{Radiative corrections and the instantaneous interaction approximation}
\label{sec:coulomb}
Let us first discuss radiative corrections to $p_\perp$-broadening. The probability for a high energy parton to acquire a transverse momentum $\k$ in the medium after a time $t$ is $\,{\cal P}(\k,t)=\int d^2\x \,e^{i\p\cdot\x} S_\x(t,0)$, 
where $S_\x(t,0)=\langle \rmTr U_\x^\dag U_\0\rangle$ is a 2-point function involving Wilson lines that encode the eikonal propagation of the high energy parton in the background field of the medium, $A^-(t,\x)$, in the amplitude and the complexe conjugate amplitudes. \footnote{The time variable  $t\equiv x^+=(x_0-x_3)/\sqrt{2}$ denotes the light-cone time. } In the approximation of independent multiple interactions one can model the medium with a classical random field correlated at equal times, i.e., $\langle A^-(t)A^-(t')\rangle \propto n\, \delta(t-t')$. \footnote{Recall that this correlator encodes the leading-order differential elastic cross section.} Hence, the single instantaneous elastic interaction exponentiates as follows, $S_\x(t,0)=1-\frac{1}{2}nC_R\,\int_0^t dt' \sigma(\x)\,S_\x(t',0)$\,, where $nC_R\,\sigma(\x)\approx \x^2\, \hat q\, /2$ is the dipole cross section, which is proportional to the quenching parameter.  The broadening probability obeys then a Fokker-Planck equation and reads ${\cal P}(\k,t)=4\pi/(\hat q\, t) \exp[-\k^2/(\hat q\, t)]$.

We turn now to the first radiative correction to $S_\x(t,0)$. It involves the emission of a soft gluon with energy $\omega$ at time $t'$ and absorbed at time $t'+\tau$ by the fast parton. We obtain the form 
\beq\label{DS}
\Delta S_\x(t,0) = \alpha_s\int^t_{0}dt' \int_{0}^{t'} d\tau \int  \frac{d\omega}{\omega^3} S_\x(t,t'+\tau)\, K(\tau,\x) \,S_\x(t',0)\,.
\eeq

where the function $K$ is given in \cite{Blaizot:2014bha}. Let us have a close look at the structure of Eq. (\ref{DS}).
Contrary to the instantaneous Coulomb interaction, the gluon fluctuation has a finite duration $\tau$. However, the integration over  $\tau$ diverges logarithmically at $0$ as we shall see. For the time being we simply cut-off the divergence at $\tau_0$. This divergence shows that radiative corrections are logarithmically sensitive to short distance scales, which were ignored in our instantaneous interaction model. 
We can then neglect the dependence on $\tau$ before and after the gluon fluctuation thanks to the logarithmic divergence, in other words the fluctuation is seen ``as is if'' it were instantaneous. Indeed, in this approximation one can assume $\tau\ll t,t'$ and integrate $\tau$ up $\tau_\text{max}\sim t'$. Therefore, the radiative correction can be ``effectively" interpreted as local and thus can be absorbed in a redefinition of the dipole cross section. We finally get,
\beq
\Delta S_\x(t,0) \simeq  \frac{1}{2}nC_R\int^t_{0}dt' S_\x(t,t') \,\Delta\sigma (\tau_\text{max},\x) \,S_\x(t',0)\,
\eeq
where $n\,C_R \,\Delta\sigma(\tau_\text{max},\x)= \alpha_s\int^{\tau_\text{max}}_{\tau_0}  d\tau \int \frac{d \omega}{\omega^3}\, K(\tau,\x) \approx \x^2 \Delta\hat q(\tau_\text{max},\x^{-2})/2$ and (switching from $(\omega,\tau)$ to $(\tau,\q)$ variables)
\beq\label{deltaqhat}
\Delta \hat q (\tau_\text{max},\k^2)=\bar\alpha \int_{\tau_0}^{\tau_\text{max}} \frac{d\tau}{\tau}\int^{\k^2}_{\hat q \tau} \frac{d\q^2}{\q^2}\,{\hat q}(\q)\,.
\eeq 
Where the dominante contribution is double logarithmic and corresponds to radiations that have been completed after a time $\tau$ by a single hard scattering with momentum squared $\q^2$ that is much larger than the scale $\hat q\,\tau$ where multiple scatterings are important. To this accuracy, we can set $\tau_\text{max}\sim t$ and $\k^2\sim \hat q\, t$ in Eq. (\ref{deltaqhat}). Finally, assuming $\hat q $ to be constant we perform the double logarithmic integrals and get for the renormalized $\hat q_\text{R}(t)\simeq \hat q +\Delta \hat q(t)$, 
\beq\label{Rqhat-1}
\hat q_\text{R}(t)\simeq \hat q \left(1+\frac{\alpha_s N_c}{2\pi}\ln^2\frac{t}{\tau_0}\right)\,.
\eeq
Integrating Eq. (\ref{Rqhat-1}) over $t$ up to $L$, we recover to double logarithmic accuracy the result by Liou, Mueller and Wu for the typical transverse momentum squared \cite{Liou:2013qya}. 
  
We shall argue that this single radiative correction will exponentiate similarly to multiple scatterings. Since the integral over $\tau$ extends to $\tau_\text{max}\sim t$, multiple radiative corrections associated with multiple scatterings could overlap. However, thanks to the logarithmic integrals the overlaps of the time integrals are down by at least a log compared to the independent radiation case \cite{Blaizot:2014bha}. So the result  yields the exponentiation of the double logs. Multiple radiative corrections along the propagation in the medium are thus accounted for by the substitution $\sigma \to \sigma_\text{R}\equiv\sigma+\Delta\sigma$ in $S(\x,t)$.  
Hence, the broadening probability yields 
\beq\label{P-mom-FP}
{\cal P}(\k,t) =\frac{4\pi}{\hat q_\text{R}(t)\,t}\, \exp\left[-\frac{\k^2}{\hat q_\text{R}(t)\,t}\right]\,,
\eeq 

In order to discuss the universality of the renormalization of the jet quenching parameter we shall consider now the  generalization of our result to the BDMPS radiative spectrum. The medium induced radiative spectrum (\ref{BDMPS-1}) is derived form following formula (for an infinite and homogeneous medium)\cite{Mehtar-Tani:2013pia}
\beq\label{BDMPS}
\omega \frac{dN}{d\omega dL} =  \frac{\alpha_sC_R}{\omega^2}\, 2\text{Re}\int_0^\infty dt \frac{\del}{\del\x}\cdot \frac{\del}{\del\y}\, {\cal K}(\x,t|\y,0)\Big|_{\x=\y=\0}\,,\nn
\eeq
where the Green's function ${\cal K}$ is related to a 3-point function (see Ref. \cite{Blaizot:2014bha} for details) that involves the eikonal propagation of the fast parton in the amplitude and the complex conjugate amplitude which is described by the Wilson lines, and the non-eikonal gluon propagation. It obeys the Dyson-like equation that accounts sequentially for multiple instantaneous interactions,
\beq\label{K}
{\cal K}(\x,t|\y,0)\,=\,{\cal K}_0(\x,t|\y,0)\,+\,\frac{1}{2}nN_c\int^t_{0} dt'\int d^2\r\,{\cal K}_0(\x,t|\r,t') \,\sigma(\r)\,{\cal K}(\r,t'|\y,0)\,.
\eeq
where ${\cal K}_0$ is the free propagator. Similarly to the 2-point function discussed previously note that the interaction with the medium is local and encoded in $\sigma(\r)$ where $\r$ is the transverse size of the dipole made by the radiated gluon and the fast emitter at time $t'$. It can be shown that radiative corrections to $\cal K$ exhibit the same logarithmic divergence when the duration of the additional gluon fluctuation $\tau\to 0$ which can occur anywhere in the time interval $t$ which is typically of the order of the BDMPS formation time  $\tau_\text{form}$. It is accounted for to double logarithmic accuracy by shifting the dipole cross-section $\sigma(\r) \to \sigma(\r) +\Delta \sigma(\tau_\text{max},\r)$, with $\tau_\text{max}\sim \tau_\text{form}$ in Eq. (\ref{K}).
This amounts to redefining the quenching parameter in the BDMPS spectrum with its renormalized value, $\hat q+\Delta\hat q(\tau_\text{form},k^2_\text{form})$, in Eq. (\ref{BDMPS}). We are able now to compute the correction to the mean energy loss. Since gluons with formation times of the order of the length of the medium dominate the integral over $\omega$, we shall set, as for $p_\perp$-broadening, $\tau_\text{form}\sim L$ and  $k^2_\text{form}\sim \hat q L$. Therefore, we readily obtain to double logarithmic accuracy, 
\be
\langle \omega \rangle \sim  \hat qL^2 \left (1+\frac{\alpha_sN_c}{2\pi}  \ln^2\frac{L}{\tau_0}\right)\,.
\ee

\section{Renormalization of the quenching parameter  }
\label{sec:RGqhat}
For large media, as soon as  $\alpha_s \ln^2(L/\tau_0)\sim1$ one has to resum the DL power corrections. This resummation of the leading logarithms involves large separation of scales. The structure of the first double logarithmic corrections is based on the strong ordering of the transverse momentum of the fluctuation $\q\ll \k$ and its duration $\tau\ll L$. Hence, the fast parton cannot resolve this fluctuation from the elastic scatterings with the medium. This being set, next corrections that yield DL's will follow the same systematics, with successive gluonic fluctuations ordered in formation time, $\tau_0 \ll \tau _1\ll ...\ll\tau_n\equiv \tau_{\text{max}} $ and in transverse momentum $m_D  \ll \q _1\ll ...\ll\q_n\equiv \k $ (or in transverse size $\x_0 \gg \x _1\gg ...\gg \x_n\equiv \x_{\text{max}} $). The difference with the standard Double-Logarithmic Approximation (DLA) is the limits of the logarithmic phase-space set by the LPM effect, i.e., multiple-scatterings since in the DLA only a single scattering contributes, which imposes that the formation time of a fluctuation to be smaller than the BDMPS formation time. In terms of our variables  $\k \gg  \hat q\,\tau $. The following equation resums the  DL corrections to all orders
\beq\label{qhat-evol}
\frac{\del \hat q (\tau,\k^2)}{\del \log \tau} = \, \int_{\hat q  \tau }^{\k^2}\frac{d\q^2}{\q^2}\, \bar \alpha(\q)\, \hat q (\tau,\q^2) \,,
\eeq
given an initial condition $ \hat q (\tau_0,\k^2)$. Here $\bar\alpha\equiv \alpha_sN_c/\pi$ and we have let the coupling running at the transverse scale $\q$.
Note again the lower limit of the $\q$ integration which accounts for the boundary between single-scattering and multiple-scatterings. The important feature of this equation is that it predicts the evolution of the jet-quenching parameter from an initial condition $\hat q_0$ (which can be computed e.g. on the lattice, or  to leading order in $\alpha_s$ as given by Eq. (\ref{qhat})). The $\tau_0$ cut-off that was introduced  to cut the logarithmic divergence in the radiative corrections, can be seen as a factorization scale.  In the case where $\hat q(\tau_0)$ is constant and for a final $\tau=L$ and $\k^2=\hat q(\tau_0)L$, merging the 2 independent variables at the end of the evolution, Eq. (\ref{qhat-evol}) can be solved iteratively \cite{Liou:2013qya}. We find then that for large $L$, the quenching parameter scales like $L^\gamma$, with the anomalous 
\be\label{anomalous-d}
\gamma=2\sqrt{\frac{\alpha_sN_c}{\pi}}\,.
\ee
Interestingly, the resummation of large DL's modifies the $L^2$ scaling of the mean energy loss, $\langle \omega \rangle \sim L^{2+\gamma}$, and seems to fall between the standard small coupling result, $\langle \omega \rangle \sim L^2 $ and the strong coupling result obtained in ${\cal N}=4$ SYM theory , $\langle \omega \rangle \sim L^3 $ \cite{Hatta:2007cs}. These results have been rederived recently from the point of view of evolution equations \cite{Iancu:2014kga}.

\section*{Acknowledgments}
This research is supported by the European Research Council under the Advanced Investigator Grant ERC-AD-267258.


\end{document}